# Can We Find the Code? An Empirical Study of Google Scholar's Code Retrieval

Shi-Shun Chen

*Abstract*—Academic codes associated with research papers are valuable resources for scholars. In specialized fields outside computer science, code availability is often limited, making effective code retrieval essential. Google Scholar is a crucial academic search tool. If a code published in the paper is not retrievable via Google Scholar, its accessibility and impact are significantly reduced. This study takes the term "accelerated degradation" combined with "reliability" as an example, and finds that, for papers published by Elsevier, only GitHub links included in abstracts are comprehensively retrieved by Google Scholar. When such links appear within the main body of a paper, even in the "Data Availability" section, they may be ignored and become unsearchable. These findings highlight the importance of strategically placing GitHub links in abstracts to enhance code discoverability on Google Scholar.

## I. INTRODUCTION

Open-source code plays a vital role in advancing academic research [1] and increasing citation impact [2]. However, identifying research papers that provide public access to code remains a significant challenge [3, 4], particularly outside the computer science domain. Platforms like *Papers with Code* (https://paperswithcode.com/) and *Code Ocean* (https://codeocean.com/) help address this issue, but they primarily serve the computer science community. Scholars in other disciplines often rely on general retrieval platforms, especially *Google Scholar* (https://scholar.google.com/), to find relevant code. Therefore, the effectiveness of Google Scholar's code retrieval is crucial. However, the code in a recently published paper by the author in Elsevier cannot be retrieved by Google Scholar [5]. This finding motivates the present study, which further finds this problem to be widespread.

In this paper, inspired by the title of [5], we take the terms "accelerated degradation" and "reliability" as an example. We first use these terms in *Google* (https://www.google.com) to find relevant papers published in *Elsevier* (https://www.sciencedirect.com/). Then, we check if these studies can be retrieved in Google Scholar with their corresponding code link.

## II. METHODS AND RESULTS

### A. Relevant Paper Determination

We used quotation marks to ensure that the retrieved papers contained the given terms. Additionally, by specifying "site:https://www.sciencedirect.com," we restricted the search to papers published by Elsevier. To find articles featuring code, we focused on GitHub (https://github.com/), the largest code-sharing platform, and included "github" as a search keyword. The searching results are shown in Fig. 1.

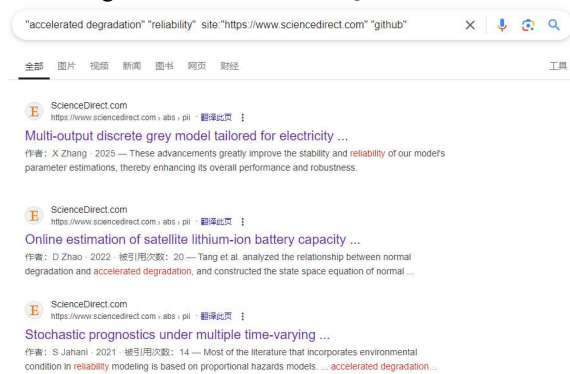

Fig. 1 Searching results in Google with the given terms.

Then, through manual selection, 25 articles that included all the terms were identified. The publication date of each article and the position of the code are listed in Table 1.

### B. Search Strategies and Results

After identifying the target articles, we conducted searches on Google Scholar using the first five words of each article's title combined with the term "GitHub". This approach was chosen to prevent exact matches that would occur if the full title were used. Additionally, we also used the exact code link from each article for comparison to minimize potential disturbances. The results are summarized in Table 2.

To have a better understanding of our search strategy, we take the search results of [6-8] as examples, which are shown in Figs 2-4, respectively.

Shi-Shun Chen is with the School of Reliability and Systems Engineering, Beihang University, Beijing 100191, China (e-mail: css1107@buaa.edu.cn).

Table 1  Publication date and code position of each paper

| Papers | Published Year | Abstract | Main body | Data/Code availability | Reference | Footnote |
|---|---|---|---|---|---|---|
| [6] | 2025 | | | √ | | |
| [9] | 2025 | | | √ | √ | |
| [10] | 2025 | | √ | | | |
| [7] | 2025 | | | √ | | |
| [11] | 2024 | | | | √ | |
| [12] | 2024 | √ | √ | | | |
| [13] | 2024 | | | | √ | |
| [14] | 2024 | | √ | √ | | |
| [15] | 2024 | | √ | √ | | |
| [16] | 2023 | | √ | | | |
| [17] | 2023 | | √ | | √ | |
| [18] | 2023 | | √ | | √ | √ |
| [19] | 2023 | | | | √ | |
| [20] | 2022 | | | | | √ |
| [21] | 2022 | | | | √ | |
| [22] | 2022 | | √ | | | |
| [23] | 2022 | | | √ | | |
| [24] | 2022 | | | √ | | |
| [25] | 2022 | | | | √ | |
| [26] | 2022 | | | √ | | |
| [27] | 2021 | | | √ | √ | |
| [8] | 2021 | | | √ | | |
| [28] | 2021 | | √ | | | |
| [29] | 2020 | | | √ | | |
| [30] | 2020 | | √ | | | |

Table 2  Search results of each article in Google Scholar

| Papers | Google Scholar search result with "github" | Google Scholar search result with exact code link |
|---|---|---|
| [6] | × | × |
| [9] | × | × |
| [10] | √ | √ |
| [7] | × | × |
| [11] | × | × |
| [12] | √ | √ |
| [13] | √ | √ |
| [14] | × | × |
| [15] | √ | √ |
| [16] | √ | √ |
| [17] | √ | √ |
| [18] | √ | √ |
| [19] | × | × |
| [20] | × | × |
| [21] | × | × |
| [22] | √ | √ |
| [23] | × | × |
| [24] | × | × |
| [25] | √ | √ |
| [26] | √ | √ |
| [27] | √ | √ |
| [8] | × | × |
| [28] | × | × |
| [29] | √ | √ |
| [30] | √ | √ |

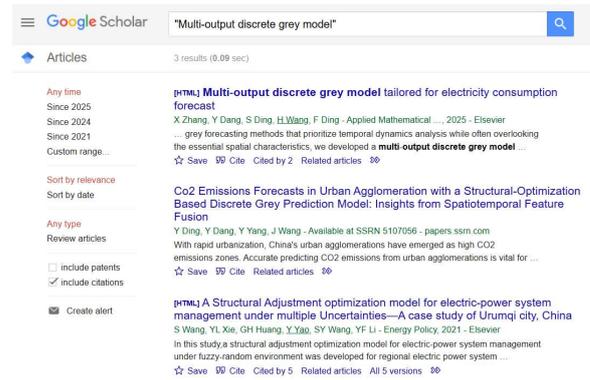

(a)

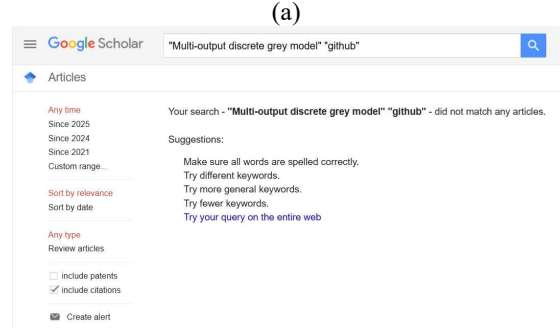

(b)

Data availability

Data and code to this article can be found in the following link: https://github.com/nuaadoctorz/SEMDGM-model.git .

(c)

Fig. 2  Search results of [6]: (a) solely by the first five words of title (b) title with "github" (c) the code link in the paper.

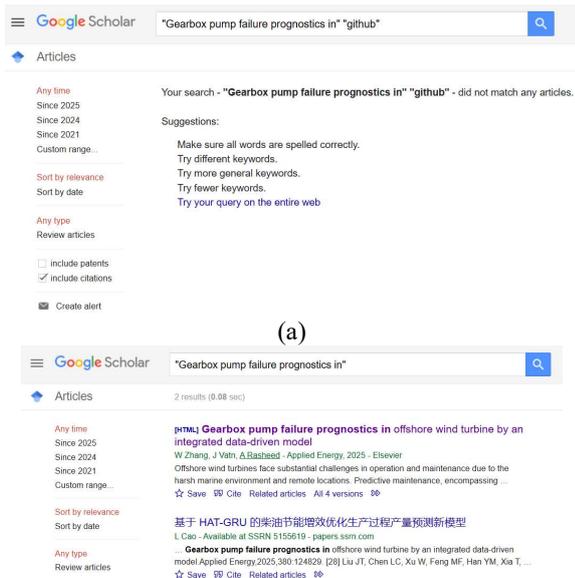
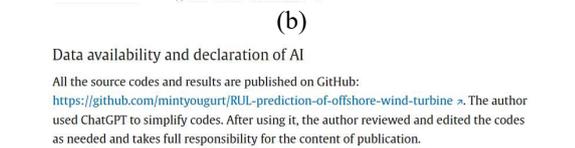
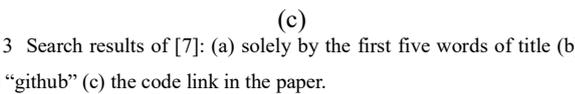

(c)

Fig. 3  Search results of [7]: (a) solely by the first five words of title (b) title with "github" (c) the code link in the paper.

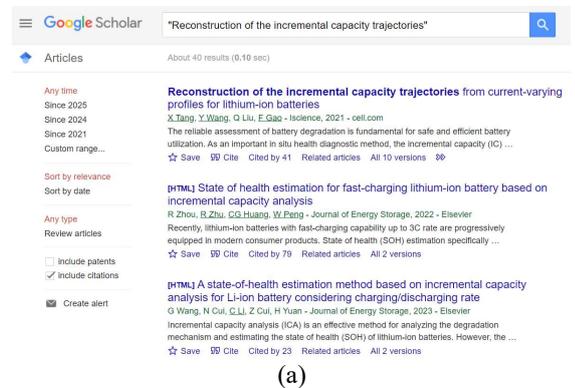
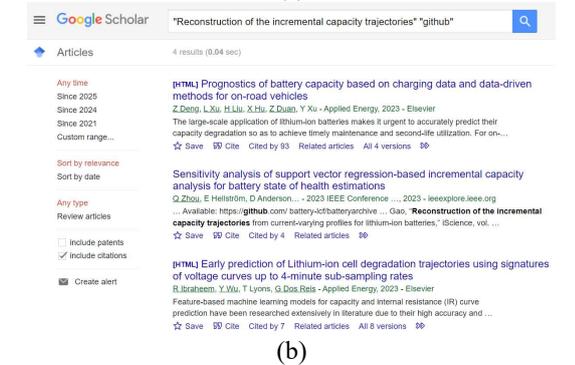

(b)

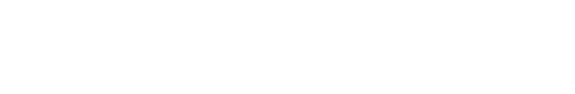

(c)

Fig. 4  Search results of [8]: (a) solely by the first five words of title (b) title with "github" (c) the code link in the paper.

From Tables 1 and 2, it can be seen that embedding code links in different sections has different impacts on its discoverability in Google Scholar:

- **Abstract**: When the abstract contained code ($n = 1$), where $n$ represents the sample size, the search success rate was 100 %. In contrast, for papers without code in the abstract ($n = 24$), the success rate dropped to 50%. While this suggests that including code in the abstract may enhance indexing accuracy, the extremely small sample size limits the reliability of this conclusion.
- **Main body**: Papers with code in the main text ($n = 10$) had a search success rate of 80 %, whereas those without code ($n = 15$) had a significantly lower success rate of 33.33 %—a difference of 46.67 %. Given the relatively balanced sample sizes, this finding is more robust and indicates that embedding code within the main text can improve Google Scholar's indexing accuracy.
- **Data/Code Availability**: Surprisingly, papers that included code in this section ($n = 11$) had a lower search success rate (36.36%) than those that did not (64.29 %, $n = 14$)—a difference of -27.92 %. This suggests that merely mentioning code in the "Data/Code Availability" section may not enhance indexing accuracy and could even negatively affect discoverability.
- **Reference**: Papers that cited code in their references ($n = 9$) had a search success rate of 55.56 %, compared to 50% for papers without such references ($n=16$)—a minimal difference of 5.56 %. This indicates that mentioning code in the references has little to no impact on indexing accuracy.
- **Footnote**: The presence of code in footnotes ($n = 2$) resulted in a 50% search success rate, while papers without footnotes ($n = 23$) had a slightly higher rate of 52.17 %—a negligible difference of -2.17 %. Given the very small sample size, this result lacks statistical reliability, and a larger dataset is needed to confirm any potential trend.

Overall, our findings suggest that embedding code in the main text and possibly the abstract may improve Google Scholar's indexing accuracy, while merely listing code in the "Data/Code Availability" section or references appears less effective.

*C. Discussions*

To further investigate the impact of code link placement on Google Scholar retrieval, we searched for "https://github.com", restricting results to Elsevier publications and setting the timeframe to 2025. Surprisingly, almost all retrieved papers included code links in their abstracts. This finding highlights the importance of including code links in abstracts to enhance discoverability.

Additionally, one highly relevant paper published by the author [5] was not retrieved even by Google. This suggests that including code links in articles may not always be sufficient to ensure code visibility on the internet.

## III. MY OWN STUDY WITH CODE LINK

Here, some papers with code link published by the author are given, because these code links are not retrieved by Google Scholar. As soon as the retrieval is recovered, this content will be deleted.

1. Title: Comparison of global sensitivity analysis methods for a fire spread model with a segmented characteristic [31]

Abstract with code link: Global sensitivity analysis (GSA) can provide rich information for controlling output uncertainty. In practical applications, segmented models are commonly used to describe an abrupt model change. For segmented models, the complicated uncertainty propagation during the transition region may lead to different importance rankings of different GSA methods. If an unsuitable GSA method is applied, misleading results will be obtained, resulting in suboptimal or even wrong decisions. In this paper, four GSA indices, i.e., Sobol index, mutual information, delta index and PAWN index, are applied for a segmented fire spread model (Dry Eucalypt). The results show that four GSA indices give different importance rankings during the transition region since segmented characteristics affect different GSA indices in different ways. We suggest that analysts should rely on the results of different GSA indices according to their practical purpose, especially when making decisions for segmented models during the transition region. All of codes are available in GitHub: https://github.com/dirge1/GSA_segmented.

2. Title: Reliability modeling and statistical analysis of accelerated degradation process with memory effects and unit-to-unit variability [5]

Abstract with code link: A reasonable description of the degradation process is essential for credible reliability assessment in accelerated degradation testing. Existing methods usually use Markovian stochastic processes to describe the degradation process. However, degradation processes of some products are non-Markovian due to the interaction with environments. Misinterpretation of the degradation pattern may lead to biased reliability evaluations. Besides, owing to the differences in materials and manufacturing processes, products from the same population exhibit diverse degradation paths, further increasing the difficulty of accurately reliability estimation. To address the above issues, this paper proposes an accelerated degradation model incorporating memory effects and unit-to-unit variability. The memory effect in the degradation process is captured by the fractional Brownian motion, which reflects the non-Markovian characteristic of degradation. The unit-to-unit variability is considered in the acceleration model to describe diverse degradation paths. Then, lifetime and reliability under normal operating conditions are presented. Furthermore, to give an accurate estimation of the memory effect, a new statistical analysis method based on the expectation maximization algorithm is devised. The effectiveness of the proposed method is verified by a simulation case and a real-world tuner reliability analysis case. The simulation case shows that the estimation of the memory effect obtained by the proposed statistical analysis method is much more accurate than the traditional one. Moreover, ignoring unit-to-unit variability can lead to a highly biased estimation of the memory effect and reliability. All of codes are available in GitHub: https://github.com/dirge1/FBM_ADT.

## IV. CONCLUSIONS

This study investigated the impact of embedding GitHub code links in different sections of research papers published by Elsevier on their discoverability in Google Scholar. By analyzing a set of papers retrieved using the terms "accelerated degradation" and "reliability", we found that the placement of code links significantly affects their indexing success. Based on the studies, several conclusions can be drawn:

- Embedding code links directly in the main text improves discoverability, whereas placing them in the Data/Code Availability section does not enhance indexing and may even reduce visibility. Additionally, including code links in references or footnotes has little to no impact on retrieval success.
- Nearly all Elsevier papers retrieved with the search term "https://github.com" for 2025 contained code links in their abstracts. This reinforces the importance of placing GitHub links in abstracts to maximize visibility.
- For scholars seeking to enhance the visibility and accessibility of their research code, we strongly recommend strategically embedding GitHub links in abstracts and main text rather than relying on dedicated sections like Data Availability.

These findings provide practical guidance for improving open-source code accessibility in academic publications. Future studies should expand the dataset across different disciplines and publishers to verify the generalizability of our findings. Additionally, analyzing Google Scholar's retrieval mechanisms and assessing the long-term effects of strategic code placement on citation impact would be valuable directions for future research.